\documentclass[twocolumn]{aastex63}
\usepackage{graphicx}	
\usepackage{amsmath}	
\usepackage{amssymb}	
\usepackage{color}
\usepackage{array}
\usepackage{lineno}
\usepackage{bookmark}
\usepackage{subfigure}
\usepackage{graphicx}
\usepackage{booktabs}
\usepackage{threeparttable}

\begin{document}

\title{Polarization in early optical afterglows of gamma-ray bursts driven by precessing jets}

\correspondingauthor{Tong Liu}
\email{tongliu@xmu.edu.cn}

\author{Bao-Quan Huang}
\affiliation{Department of Astronomy, Xiamen University, Xiamen, Fujian 361005, China}

\author[0000-0001-8678-6291]{Tong Liu}
\affiliation{Department of Astronomy, Xiamen University, Xiamen, Fujian 361005, China}

\begin{abstract}
Jet precessions are widely involved in astrophysical phenomena from galaxies to X-ray binaries and gamma-ray bursts (GRBs). Polarization presents a unique probe of the magnetic fields in GRB jets. The precession of GRBs relativistic jets will change the geometry within the observable emitting region of the jet, which can potentially affect the polarization of the afterglow. In this paper, we take into account jet precession to study the polarization evolution and corresponding light curves in GRB early optical afterglows with ordered and random magnetic field geometries. We find that the jet precession in long-lived engines can significantly reduce the polarization degree (PD) regardless of the magnetic field structure. The strongest PD attenuation is found when the line of sight is aligned with the precession axis. Our results show that jet precession can provide new insight into the low PD measured in the early optical afterglows of GRBs.
\end{abstract}


\section{Introduction}
The properties of precessing jets have been widely studied in galaxies \citep[see e.g.,][]{Miley1980,Begelman1984,Lu1990,Proctor2011}, X-ray binaries such as SS 433 \citep[e.g.,][]{Margon1984}, Swift J164449.3+573451 \citep{Burrows2011}, and MAXI J1820+070 \citep{Ma2021}, as well as in gamma-ray bursts \citep[GRBs, see e.g.,][]{Blackman1996,Portegies1999,Reynoso2006,Liu2010,Sun2012,Hou2014,Huang2019,Huang2021}.

Being the strongest explosions with $\gamma$-ray emission in the Universe, GRBs are believed to be powered by ultra-relativistic jets \citep[see a review by][]{zhang2018book}. GRBs have two types of progenitors, i.e., the coalescence of compact objects and the collapse of massive stars, which correspond to short- and long-duration GRBs (SGRBs and LGRBs), respectively \citep[see the reviews by][]{Woosley2006,Nakar2007,Liu2017a}. If misalignment in the directions of the angular momenta of merged compact objects or anisotropic collapsars occurs, the new-born compact objects, involving a massive millisecond pulsar \citep[e.g.,][]{Usov1992,Dai1998a,Dai1998b,Zhang2001,Li2016,Hou2021} and a rapidly rotating and hyperaccreting black hole \citep[for reviews, see][]{Liu2017a}, will power the precessing jets \citep{Liu2017b}. Additionally, precession has been proposed as one of the probable scenarios powering quasi-periodic oscillations during the GRB gamma-ray prompt emission \citep{Tarnopolski2021}.

The multiwavelength afterglow that is usually detected after the short-lived prompt $\gamma$-ray emission originates from the synchrotron radiation of electrons accelerated by forward and reverse shocks (FS and RS, respectively), produced when the jets interact with the circumburst medium \citep[e.g.,][and references therein]{Zhang2006}. In this context, the polarimetry of GRB afterglows has become an important topic as it can provide crucial information on the properties of the magnetic fields embedded in the jets \citep[for a review, see][]{Gill2021}. In recent years, a faster response of the GRB detection technology has allowed early-time polarization measurements of several afterglows \citep[e.g.,][]{Gorbovskoy2016,Laskar2019,Buckley2021}. Currently, the measurements for early optical afterglows show a polarization degree (PD) with a wide range of $1\%\sim30\%$ \citep[e.g.,][]{Covino2016,Steele2017}. Different origins have been proposed regarding the PD in this range. The dust from the Galactic interstellar medium or GRB host galaxies \citep[e.g.,][]{Jordana2020} and the magnetic fields produced in the FSs \citep[e.g.,][]{King2014,Jordana2021} might result in the low PD. To account for the higher PD ($>10\%$) measured in RSs, large-scale ordered magnetic fields are thought to be the primary origin, such as GRB 090102 \citep{Steele2009}, GRB 110205A \citep{Steele2017}, and GRB 120308A \citep{Mundell2013}.

It is expected that the large-scale ordered magnetic fields inherited from the central engine remain in the jets \citep[e.g.,][]{Spruit2001,Lazzati2006}, and, hence, a high PD (up to $\sim 60 \%$) has been predicted to appear in the early phase of the afterglows \citep{Granot2003}. However, compared with the maximum theoretically predicted value, most of the measured PD of the early optical afterglows whose jets are suggested to possess large-scale ordered magnetic fields are still substantially low \citep{Rybicki1979,Lyutikov2003}. This means that there are other factors that weaken the PD even if large-scale ordered magnetic fields robustly exist \citep{Zhang2011,Deng2017}. For instance, the RSs can be significantly suppressed due to the high magnetization in the jet ejecta \citep{Zhang2005,Giannios2008}, and then, high PD is not expected for the early afterglow emission, or a mixed magnetic field structure exists in the jets \citep{Lan2019a}. These factors, in principle, can exist individually or simultaneously. In addition to the magnetic field geometry, the jet structure can also affect the PD \citep{Rossi2004}. Therefore, more scenarios, such as jet precession, should be taken into account to understand the low PD measured in the early optical afterglows of GRBs.

In this paper, we aim to study the effects of jet precession on the polarization evolution of the GRB early afterglows. The paper is arranged as follows: We introduce the jet precession model in Section 2 and the polarization of synchrotron emission in an FS-RS system is calculated in Section 3. The main results are given in Section 4, and a brief summary is given in Section 5.

\section{Dynamics}

We employ the treatment used in \cite{Huang2021} to calculate the afterglow polarization of GRBs produced by the long-lived precessing jets. In this framework, the precession process causes the observer to view an intermittent jet, which we model as a series of subjets.

In the first precession period, owing to the interaction between the subjets and the interstellar medium (ISM, considered here), a pair of relativistic shocks are generated for each subjet, i.e., an FS propagating into the ISM and a RS propagating into the subjet. There are four regions separated by this pair of shocks with a contact discontinuity, including the unshocked ISM (Region 1), the shocked ISM (Region 2), the shocked subjet (Region 3), and the unshocked subjet (Region 4). Among them, Regions 2 and 3 are treated as a system with a uniform Lorentz factor, which decelerates after sweeping in enough ISM. In order for the jet precession to have a significant impact on the overall observed polarization, the active timescale of the central engine needs to be far larger than the precession period, i.e., an energy injection/long-lived engine scenario \citep{Laskar2015}. Consequently, the subsequent subjets in each period catch up with the previous system, leading to energy injection and refreshing of the RS. With increasing ISM mass, the velocity of the system reduces more quickly, hence less time is spent by later subjets to catch up.

\subsection{Dynamics before RS crossing time}

For a long-lasting and magnetized central engine, an extended version of the mechanical model proposed by \cite{Beloborodov2006} is employed here to delineate the evolution of the Lorentz factor $\Gamma$ of the system before the time when the RS passes through the subjet in the last period (i.e., the RS crossing time), which can be obtained by solving the differential equations listed as follows \citep[e.g.,][]{Ai2021}:
\begin{equation}
\label{mass}
\frac{\beta}{r^{2}_{\rm d}} \frac{d}{dr_{\rm d}} (r^{2}_{\rm d} \Sigma \Gamma) - \Gamma [\rho_{\rm r}(\beta - \beta_{\rm r}) + \rho_{\rm f}(\beta_{\rm f} - \beta)] = 0,
\end{equation}
\begin{equation}
\label{momentum}
\begin{split}
\frac{\beta}{r^{2}_{\rm d}} & \frac{d}{dr_{\rm d}} (r^{2}_{\rm d} \Gamma^{2} H \beta) - \Gamma^{2} \beta [h_{\rm r} (\beta-\beta_{\rm r}) + h_{\rm f} (\beta_{\rm f} - \beta)] \\
& + \frac{\beta}{4\pi} \frac{d}{dr_{\rm d}} (\Gamma^{2} \beta {\cal B}) + \frac{\beta \Gamma^{2}}{4\pi} [B_{\rm r}^{2} \beta_{\rm r} -  B_{\rm f}^{2} \beta_{\rm f}] + (p_{\rm f} - p_{\rm r}) \\ & + \frac{\Gamma^{2} (1 + \beta^{2})}{8\pi} (B_{\rm f}^{2} - B_{\rm r}^{2}) + \frac{(1+\beta^{2}) \Gamma^{2}{\cal B}} {4\pi r_{\rm d}} = 0,
\end{split}
\end{equation}
\begin{equation}
\label{energy}
\begin{split}
\frac{\beta}{r^{2}_{\rm d}} & \frac{d}{dr_{\rm d}} (r^{2}_{\rm d} \Gamma^{2} H) - \Gamma^{2} [h_{\rm r} (\beta - \beta_{\rm r}) + h_{\rm f}(\beta_{\rm f} - \beta)] \\
& - \beta \frac{dP}{dr_{\rm d}} - (\beta_{\rm r} p_{\rm r} - \beta_{\rm f} p_{\rm f}) + \frac{\beta}{8\pi} \frac{d}{dr_{\rm d}} [(1+\beta^2) \Gamma^{2} {\cal B}] \\ & + \frac{(1 + \beta^{2}) \Gamma^{2}}{8\pi} (\beta_{\rm r} B_{\rm r}^{2} - \beta_{\rm f} B_{\rm f}^{2}) + \frac{\Gamma^{2} \beta}{4\pi} (B_{\rm f}^{2} - B_{\rm r}^{2}) \\ & + \frac{\beta \Gamma^2 {\cal B}}{2\pi r_{\rm d}} = 0,
\end{split}
\end{equation}
\begin{equation}
\label{dB}
\frac{d{\cal B}}{dr_{\rm d}} = \frac{1}{4\pi r_{\rm r}^{2}} \frac{d{\cal B}_{\rm sph}}{dr_{\rm d}} - 2{\cal B} \frac{r_{\rm d}}{r_{\rm r}^{2}},
\end{equation}
\begin{equation}
\label{H}
H=\Sigma c^{2} + \frac{\hat{\gamma}}{\hat{\gamma} - 1} P.
\end{equation}
Here, $\Sigma \equiv \int_{r_{\rm r}}^{r_{\rm f}}\rho \, dr$, $P \equiv \int_{r_{\rm r}}^{r_{\rm f}} p \, dr$, $H \equiv \int_{r_{\rm r}}^{r_{\rm f}} h \, dr$, and ${\cal B} \equiv \int_{r_{\rm r}}^{r_{\rm f}} B^{2} \, dr$, with $r_{\rm r}$, $r_{\rm d}$, and $r_{\rm f}$ are the distances of the RS, the contact discontinuity, and the FS from the central engine, which represent the integrated quantities of the mass density, the thermal pressure, the enthalpy, and the square of the magnetic field, respectively. $\rho_{\rm r}$ $(\rho_{\rm f})$, $p_{\rm r}$ $(p_{\rm f})$, $h_{\rm r}$ $(h_{\rm f})$, and $B_{\rm r}$ $(B_{\rm f})$ denote the values of $\rho$, $p$, $h$, and $B$ immediately behind the RS (FS) in the comoving of the system, respectively, which can be acquired by the shock jump condition \citep[e.g.,][]{Zhang2005}. $\beta_{\rm r}$, $\beta$, and $\beta_{\rm f}$ are the velocities of the RS, the contact continuity, and the FS in the lab frame, respectively. $\hat{\gamma}$ is the adiabatic index. In addition, ${\cal B}_{\rm sph}$ is the integrated quantity of the square of the magnetic field but with an integral over the volume, and its evolution with $r_{\rm d}$ can be expressed as $d{\cal B}_{\rm sph} / dr_{\rm d} = 4 \pi \sigma_{\rm r} L_{\rm ej} (\beta_{\rm ej} - \beta_{\rm r}) / \Gamma \Gamma_{\rm ej} \beta \beta_{\rm ej} c (1 + \sigma_{\rm ej})$, where $L_{\rm ej}$, $\Gamma_{\rm ej}$, $\beta_{\rm ej}$, $\sigma_{\rm ej}$, and $\sigma_{\rm r}$ are the luminosity of the central engine, the Lorentz factor of the unshocked subjets, the velocity of the unshocked subjets in the lab frame, the magnetization parameter of the unshocked subjets, and the magnetization parameter at the RS downstream, respectively.

\begin{figure}
\centering
\includegraphics[width=1.0\linewidth,height=0.97\linewidth]{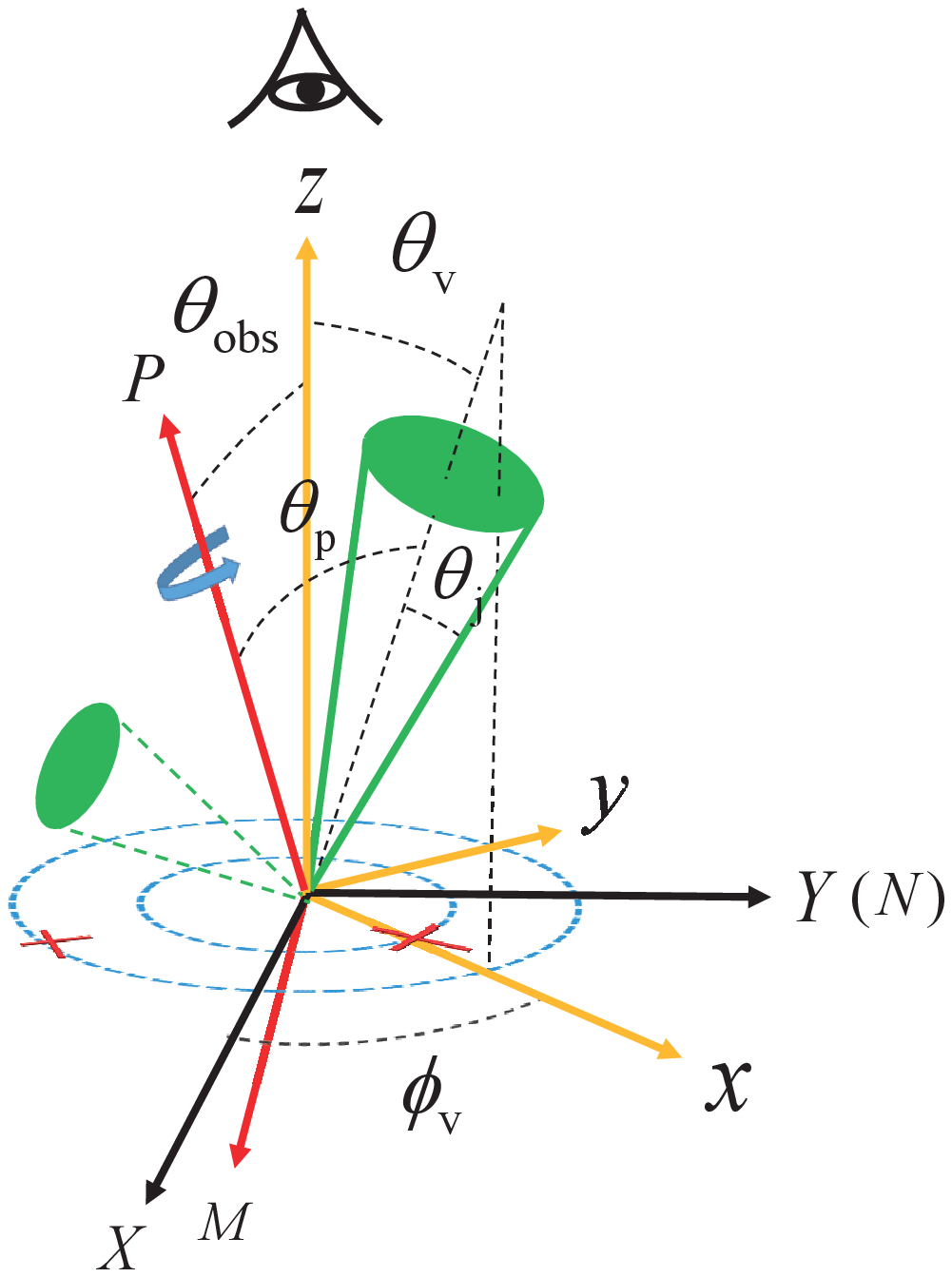}
\caption{Illustration of a GRB precessing jet.}
\label{cartoon}
\end{figure}

The mass of the system contains two parts: the mass of Regions $2$ and $3$. The mass evolution of these two regions with radius $r_{\rm d}$ can be expressed as \citep[e.g.,][]{Huang1999,Nava2013}
\begin{equation}
\label{m2}
dm_{2} = 4 \pi r_{\rm f}^{2} n_{1} m_{\rm p} dr_{\rm f},
\end{equation}
and
\begin{equation}
\label{m3}
dm_{3} = 4 \pi r_{\rm r}^{2} n_{3} m_{\rm p} \Gamma dX,
\end{equation}
respectively, where $dr_{\rm f} = \beta_{\rm f} dr_{\rm d} / \beta$, $dX = (\beta_{\rm ej} - \beta) dr_{\rm d} / (\Gamma n_{3} / \Gamma_{\rm ej} n_{4} - 1) \beta$, $n_{1}$, $n_{3}$, and $n_{4}$ are the number densities of Regions $1$, $3$, and $4$ in the frame themselves, and $m_{\rm p}$ is the proton mass. For convenience, quantities with a prime are defined in the comoving frame.

Adopting a constant Lorentz factor $\Gamma_{\rm ej}$ for the unshocked subjets, there exists a scenario where the RS has already passed through the current subjet before the next subjet catches up. For this situation, we assume that there is not enough time for Regions $2$ and $3$ to independently evolve between after the RS crosses the current subjet and before the next subjet catches up and assume that the dynamic evolution of the system during this phase is only dominated by the FS. These assumptions should be reasonable for the case considered in this work given that the precession period is much less than the active timescale of the central engine. On the other hand, we describe the dynamical evolution with the ejection time $\tau$ of the subjets from the central engine, instead of radius $r_{\rm d}$, which can be achieved by replacing $dr_{\rm d}$ with $d\tau$ in the above equations via the expression
\begin{equation}
\label{dtau}
{dr_{\rm d}}=
\left\{
\begin{array}{ll}
\displaystyle \frac{c \beta \beta_{\rm ej}}{(\beta_{\rm ej} - \beta_{\rm r})} d\tau, \quad T_{\rm start} < \tau < T_{\rm end},\\
\displaystyle \frac{c \beta \beta_{\rm ej}}{(\beta_{\rm ej} - \beta)} d\tau, \quad {\rm others},
\end{array}
\right.
\end{equation}
where $T_{\rm start} = [(i-1)\Delta t + (j-1)T]$, $T_{\rm end} = [i\Delta t + (j-1)T]$, $i$ $(=1,2,\ldots,k)$ denotes the sequence number of subjets within each period, $j$ $(=1,2,\ldots,t_{\rm end}/T)$ represents the sequence number of the precession period, $T$ is the precession period, and $\Delta t = T/k$ \citep{Huang2021}. Moreover, by combining Equation (\ref{dtau}) with formula $dr_{\rm d} = c \beta dt/(1-\beta)(1+z)$, where $t$ is the time measured in the observer frame and $z$ is the redshift, we can derive the relation between differential quantities $dt$ and $d\tau$.

\begin{table*}[t]
\footnotesize
\centering
\caption{Definitions, units, and values of the basic parameters.}
\label{lab1}
\tabcolsep 18pt %space between two columns.
\resizebox{\textwidth}{!}
{
\begin{threeparttable}
\begin{tabular}{clcc}
\toprule
Parameters           & Definitions                                             &  Unit               &   Values in Case I (II)\tnote{*}      \\
\hline
$L_{\rm ej}$         & luminosity of the central engine                    & ${\rm erg/s}$       &    $10^{49}$                   \\
$\Gamma_{\rm ej}$    & Lorentz factor of the unshocked sub-jets            &                     &    300                         \\
$\sigma_{\rm ej}$    & magnetization parameter of the unshocked sub-jets   &       &   $10^{-1}$ ($10^{-3}$)                      \\
$n_{1}$              & number density of the ISM                           & ${\rm cm^{-1}}$     &   $1$                          \\
$\varepsilon_{\rm e,2}$ & fraction of the internal energy density in the FS downstream shared by the electrons          &    &   $10^{-1}$     \\
$\varepsilon_{B,2}$  & fraction of the thermal energy density in the FS downstream shared by the magnetic field          &    &  $10^{-3}$     \\
$p_{2}$              & power-law index of the energy distribution of the shocked electrons in the FS downstream     &    &   $2.5$         \\
$\varepsilon_{\rm e,3}$ & fraction of the internal energy density in the RS downstream shared by the electrons  &    &   $10^{-1}$ ($10^{-3}$)  \\
$a$             & power-law index of the ordered magnetic field decaying at Region 3 after the RS crossing time     &   &       $1$           \\
$p_{3}$              & power-law index of the energy distribution of the shocked electrons in the RS downstream           &   &   $2.5$         \\
$T$                  & precession period of the jets                          &  ${\rm s}$          &   10                \\
$t_{\rm end}$        & age of the jets                                         &  ${\rm ks}$         &   $1$               \\
$\theta_{\rm p}$     & precession angle                                        &  degree             &   $7$              \\
$\Pi_{\rm 0}$        & linear PD for synchrotron emission from a point-like region     &     &   $0.6$  \\
$\delta$             & orientation of the aligned magnetic field                  &  rad                   &   $\pi/4$               \\
$\theta_{\rm j}$     & half-opening angle of jets                           &  degree            &   5             \\
$z$                  & redshift                            &                     &   1             \\
\bottomrule
\noalign{\smallskip}
\end{tabular}
\begin{tablenotes}
\centering
\footnotesize
\item [*] The values of the basic parameters in case II are the same as that in case I and are, therefore, not shown for brevity.
\end{tablenotes}
\end{threeparttable}
}
\end{table*}

\subsection{Dynamics after RS crossing time}

After the RS crossing time, Regions $2$ and $3$ begin to evolve independently. For Region $2$, the evolution of the Lorentz factor $\Gamma_{2}$ can be expressed as \citep[e.g.,][]{Chen2021}
\begin{equation}
\label{gamma2}
\frac{d\Gamma_{2}}{dR_2} = - \frac{\Gamma_{2}^{2} - 1}{[\epsilon + 2(1-\epsilon) \Gamma_{2}] m_{2} } \frac{dm_{2}}{dR_2},
\end{equation}
where $dm_{2}/dR_2 = 4 \pi R^{2}_2 n_{1} m_{\rm p} dR_2 $, with $R_2$ being the distance from the central source, and $\epsilon$ is the fraction of the thermal energy that is radiated, set as zero here. Likewise, for Region $3$, the dynamical evolution is described by the Blandford-McKee (BM) self-similar solution in the context of a thick shell. Because of the limitation of the BM self-similar solution, which is only valid in the ultrarelativistic and relativistic phases, our calculation is terminated before entering the nonrelativistic phase.

\section{Polarization}

We consider synchrotron radiation as the emission mechanism of the afterglows. For a given frequency $\nu'$, the synchrotron radiation power is \citep{Rybicki1979}
\begin{equation}
\label{Fv}
P'_{\rm syn}({\nu}') = \frac{\sqrt{3} e^3 B' \sin \theta'_{B}}{m_{\rm e}c^2}\int\nolimits_{\gamma_{\rm e,min}'}^{\gamma_{\rm e,max}'} \bigg(\frac{dN_{\rm e}'}{d \gamma_{\rm e}'} \bigg) F \bigg(\frac{\nu'}{\nu_{\rm c}'} \bigg) d \gamma_{\rm e}',
\end{equation}
where $e$ is the electron charge, $m_{\rm e}$ is the electron mass, and $\theta'_{B}$ is the pitch angle of the electrons, namely, the angle between the direction of the velocity of the electrons and the direction of the magnetic field $B'$. $F({\nu'}/{\nu_{\rm c}'})=(\nu'/\nu_{\rm c}') \int\nolimits_{\nu'/\nu_{\rm c}'}^{+ \infty} K_{5/3}(x) dx$, with $K_{5/3}(x)$ being a modified Bessel function of order 5/3, is the synchrotron spectrum function, where $\nu_{\rm c}'=3 e B' \gamma_{\rm e}'^{2} \sin \theta'_{B}/4 \pi m_{\rm e} c$ is the critical frequency of electrons with Lorentz factor $\gamma'_{\rm e}$. $\gamma'_{\rm e,min}$, $\gamma'_{\rm e,max}$, and $dN'_{\rm e} / d\gamma'_{\rm e}$ are the minimum, the maximum Lorentz factor, and the energy spectrum of the shock-accelerated electrons, respectively \citep[e.g.,][]{Huang2003,Huang2021}.

In the calculations of the afterglow polarization, two types of magnetic field geometries confined in the shock plane, globally ordered and random, are considered, where globally ordered magnetic fields involve both aligned and toroidal configurations. For simplicity, we assume that the random magnetic field is only generated by the FS and exists in Region $2$ and assume that the ordered magnetic field only originates from the magnetic central engine and exists in Region $3$ \citep{Sari1999}. Therefore, the random magnetic field is expressed as $B'_{2}=\sqrt{32 \pi \Gamma (\Gamma-1) n_{1} m_{\rm p} \varepsilon_{B,2}c^2}$, with $\varepsilon_{B,2}$ being the fraction of the thermal energy density in the FS downstream shared by the random magnetic field; note that $\Gamma$ is replaced by $\Gamma_2$ in the formula after the RS crossing time. Likewise, the ordered magnetic field can be solved by the shock jump condition \citep[e.g.,][]{Ai2021} before the RS crossing time. Since the evolution of the ordered magnetic field cannot be obtained from the shock jump condition after the RS crossing time, we assume that the ordered magnetic field power law decays with radius, i.e., $B'_{3}\propto R_{\rm 3}^{-a}$, with $R_3$ being the distance of Region 3 from the central source and $a = 1$ (see Table \ref{lab1}).

Considering an off-axis observer, we initially set a lab frame $xyz$ with the $z$ axis aligned with the line of sight and the $x$ axis oriented along the projected direction of the jet axis moving due to the precession in the plane of the sky, as shown in Figure \ref{cartoon}. In this frame, $\theta_{\rm v}$ denotes the angle between the line of sight and the jet axis, $\theta_{\rm j}$ represents the jet half-opening angle, and the position of a jet element is described as $(\theta, \phi)$.

\subsection{Stokes parameters in a random magnetic field}

In the random magnetic field in Region 2, the local PD of a point-like region with a random magnetic field can be expressed by \citep[e.g.,][]{Toma2009,Lan2016}
\begin{equation}
\label{eq:Pi0}
\Pi_{\rm p} = - \Pi_{\rm 0} \frac{\left< (\sin \theta_B^\prime)^{1-m} \cos(2 \phi_B^\prime) \right>}{\left< (\sin \theta_B^\prime)^{1-m} \right>},
\end{equation}
where $\cos (2 \phi'_B) = (2 \sin^2 \eta' / \sin^2 \theta_B') - 1$ and $\sin \theta'_B = (1 - {D}^2 \sin^2 \theta \cos^2 \eta')^{1/2}$, with $D = 1 / \Gamma (1-\beta \cos \theta)$ being the Doppler factor, $\Pi_{\rm 0}$ is the PD from a smaller region, $m$ is obtained from the power-law spectrum $ f'_{\nu'} \propto (\nu')^{m}$ \citep{Sari1998}, and $\langle\rangle$ represents the average over $\eta'$ from $0$ to $2\pi$.

By integrating the flux received from those point-like regions with the same observation time in Region 2, one can calculate the observed flux density as follows:
\begin{equation}
\label{Fvobs2}
F_{\nu_{\rm obs},{i},2} = \frac{1+z}{4 \pi D_L^2} \int_{\theta_{-}}^{\theta_{+}} \int_{-\Delta \phi}^{\Delta \phi} P'_{\rm syn}({\nu}') {D}^3 \sin \theta d \theta d \phi,
\end{equation}
where subscript ``2" represents integration carried out in Region $2$, $\nu' = (1+z)\nu_{\rm obs}/D$ with $\nu_{\rm obs}$ is the observed frequency, and $D_{L}$ is the luminosity distance in the standard $\Lambda$CDM cosmology model ($\Omega_M=0.27$, $\Omega_\Lambda=0.73$, and $H_0=71~\rm km~s^{-1}~Mpc^{-1}$).
Additionally, the Stokes parameters, $Q_{\nu_{\rm obs},{i},2}$ and $U_{\nu_{\rm obs},{i},2}$, are given by \citep[e.g.,][]{Geng2018}
\begin{widetext}
\begin{equation}
\label{Q2U2}
\left\{
\begin{array}{c} Q_{\nu_{\rm obs},{i},2} \\ U_{\nu_{\rm obs},{i},2}
\end{array}
\right\}
= \frac{1+z}{4 \pi D_L^2} \int_{\theta_{-}}^{\theta_{+}} \int_{-\Delta \phi}^{\Delta \phi} P'_{\rm syn}({\nu}') \Pi_{\rm p} {D}^3 \sin \theta d \theta
\left\{
\begin{array}{c} \cos(2\phi) \\ \sin(2\phi)
\end{array}
\right\} d \phi.
\end{equation}
\end{widetext}
Here, the corresponding integral limits are \citep[e.g.,][]{Wu2005,Geng2018}
\begin{equation}
\label{limit}
\Delta \phi =
\left\{
\begin{array}{l}
\displaystyle \pi \Theta (\theta_{\rm v}-\theta_{\rm j}),~~~~~~~~~~~~~~~~~~~~~~~~~\theta \le \theta_{-}, \\
\displaystyle \arccos \left( \frac{\cos \theta_{\rm j} - \cos \theta_{\rm v} \cos \theta}{\sin \theta_{\rm v} \sin \theta} \right),~~\theta_{-} < \theta < \theta_{+}, \\
\displaystyle 0,~~~~~~~~~~~~~~~~~~~~~~~~~~~~~~~~~~~~~~~~\theta \ge \theta_{+},
\end{array}
\right.
\end{equation}
where $\Theta$ is the Heaviside step function, $\theta_{-} = \left| \theta_{\rm j} - \theta_{\rm v} \right|$, and $\theta_{+} = \theta_{\rm j} + \theta_{\rm v}$. In addition, $U_{\nu_{\rm obs},{i},2}$ ($\propto \int \sin(2\phi) d \phi$) is equal to zero.

\subsection{Stokes parameters in an ordered magnetic field}

In the ordered magnetic field of Region 3, similar to that in the random magnetic field, the observed flux density is given by
\begin{equation}
\label{Fvobs3}
F_{\nu_{\rm obs},{i},3} = \frac{1+z}{4 \pi D_L^2} \int_{\theta_{-}}^{\theta_{+}} \int_{-\Delta \phi}^{\Delta \phi} P'_{\rm syn}({\nu}') {D}^3 \sin \theta d \theta d \phi,
\end{equation}
and the other two Stokes parameters are
\begin{widetext}
\begin{equation}
\label{Q3U3}
\left\{
\begin{array}{c} Q_{\nu_{\rm obs},{i},3} \\ U_{\nu_{\rm obs},{i},3}
\end{array}
\right\}
= \Pi_{\rm 0} \frac{1+z}{4 \pi D_L^2} \int_{\theta_{-}}^{\theta_{+}} \int_{-\Delta \phi}^{\Delta \phi} P'_{\rm syn}({\nu}') {D}^3 \sin \theta d \theta
\left\{
\begin{array}{c} \cos(2\chi) \\ \sin(2\chi)
\end{array}
\right\} d \phi,
\end{equation}
\end{widetext}
where subscript ``3" represents integration carried out in Region 3 and $\chi$ is the position angle of the polarization for a point-like region. Here, if the ordered magnetic field exists with an aligned configuration, the position angle can be expressed as \citep[e.g.,][]{Lan2016}
\begin{eqnarray}
\label{PAA}
\chi = \phi + \arctan \left [ \frac{\cos \theta - \beta}{\cos \theta (1- \beta \cos \theta)} \cot(\phi-\delta) \right ],
\end{eqnarray}
where $\delta$ is the direction of the aligned magnetic field. If the ordered magnetic field possesses a toroidal configuration, the position angle is given by
\begin{widetext}
\begin{eqnarray}
\label{PAT}
\chi=\phi+ \arctan \left( \frac{\cos \theta - \beta}{(1- \beta \cos \theta)} \frac{\sin \theta_{\rm v} \sin \phi}
{(\cos \theta_{\rm v} \sin \theta - \sin \theta_{\rm v} \cos \theta \cos \phi)} \right).
\end{eqnarray}
\end{widetext}
Additionally, it should be noted that $\sin (2 \chi)$ is an odd function of $\phi$ in the toroidal configuration \citep{Spruit2001}, and, hence, $U_{\nu_{\rm obs},{i},3} \propto \int_{-\Delta \phi}^{\Delta \phi} \sin (2 \chi) d \phi = 0$ will always remain in this configuration.

\subsection{Effects of jet precession}

We accumulate the Stokes parameters calculated in each subjet at the same observation time to obtain the total polarization effect. Because the Stokes parameters in each subjet are acquired in the moving coordinate system $xyz$, we transition the Stokes parameters from the moving coordinate system $xyz$ to a global coordinate system $XYz$ with the $X$ axis aligned with the projected direction of the precession axis in the plane of the sky, as shown in Figure \ref{cartoon}, which can be accomplished by the rotation matrix \citep[e.g.,][]{Lan2019b}, i.e.,
\begin{equation}
\label{RM}
\left(\begin{matrix}F^{X}_{\nu_{\rm obs},{i,j}}\\ Q^{X}_{\nu_{\rm obs},{i,j}}\\ U^{X}_{\nu_{\rm obs},{i,j}}\end{matrix}\right)
=\left(\begin{matrix}1&0&0\\ 0&\cos2\alpha&\sin2\alpha\\ 0&-\sin2\alpha&\cos2\alpha\end{matrix}\right)
\left(\begin{matrix}F_{\nu_{\rm obs},{i,j}}\\ Q_{\nu_{\rm obs},{i,j}}\\ U_{\nu_{\rm obs},{i,j}}\end{matrix}\right).
\end{equation}
Here, $\alpha = 2\pi-\phi_{\rm v}$ is the rotation angle moving anticlockwise from the $x$ axis to the $X$ axis, wherein $\phi_{\rm v}$ is the azimuthal angle of the jet axis in the coordinate system $XYz$, and its value can be solved by the following relations:
\begin{widetext}
\begin{equation}
\label{relation1}
\sin\theta_{\rm v}\cos\phi_{\rm v}=\sin\theta_{\rm obs}\cos\phi_{\rm obs}\cos\theta_{\rm p}
+\cos\theta_{\rm obs}\cos\phi_{\rm obs}\sin\theta_{\rm p}\cos\phi_{\rm p}
-\sin\phi_{\rm obs}\sin\theta_{\rm p}\sin\phi_{\rm p},
\end{equation}
\begin{equation}
\label{relation2}
\sin\theta_{\rm v}\sin\phi_{\rm v}=\sin\theta_{\rm obs}\sin\phi_{\rm obs}\cos\theta_{\rm p}
+\cos\theta_{\rm obs}\sin\phi_{\rm obs}\sin\theta_{\rm p}\cos\phi_{\rm p}
+\cos\phi_{\rm obs}\sin\theta_{\rm p}\sin\phi_{\rm p},
\end{equation}
\begin{equation}
\label{eq:relation3}
\cos\theta_{\rm v}=\cos\theta_{\rm obs}\cos\theta_{\rm p}-\sin
\theta_{\rm obs}\sin\theta_{\rm p}\cos\phi_{\rm p},	
\end{equation}
\end{widetext}
where ($\theta_{\rm obs}, \phi_{\rm obs}$) represents the coordinate of the precession axis in the coordinate system $XYz$ and ($\theta_{\rm p}, \phi_{\rm p}$) denotes the coordinate of the jet axis in the coordinate system $MNP$ with the $M$ axis on the $zX$ plane and $P$ the precession axis, as shown in Figure \ref{cartoon}.

After coordinate transformation, the PD and polarization angle (PA) of the calculated radiation involving the effect of jet precession can be expressed as
\begin{equation}
\label{PD}
\Pi^{X} = \frac{\sqrt{{Q^{X}_{\nu_{\rm obs}}}^2 + {U^{X}_{\nu_{\rm obs}}}^2}}{F_{\nu_{\rm obs}}^{X}},
\end{equation}
and
\begin{equation}
\label{PA}
\chi^{X} = \frac{1}{2}\arctan \left(\frac{U_{\nu_{\rm obs}}^{X}}{Q_{\nu_{\rm obs}}^{X}} \right),
\end{equation}
respectively, where $F_{\nu_{\rm obs}}^{X} = \sum_{i} \sum_{j} F^{X}_{\nu_{\rm obs},i,j}$, $Q_{\nu_{\rm obs}}^{X} = \sum_{i} \sum_{j} Q^{X}_{\nu_{\rm obs}, i,j}$, and $U_{\nu_{\rm obs}}^{X} = \sum_{i} \sum_{j} U^{X}_{\nu_{\rm obs},i,j}$, with $j\; (=2,3)$ representing the radiating regions.

Notably, the actual value of the PA cannot be acquired only by Equation (\ref{PA}). To determine the real value of the PA, the sign of the Stokes parameters $U_{\nu_{\rm obs}}^{X}$ and $Q_{\nu_{\rm obs}}^{X}$ also need to be considered here. Namely, the real value is equal to $\chi^{\rm X}$ if $Q_{\nu_{\rm obs}}^{X} > 0$, to $\chi^{X}+\pi/2$ if $Q_{\nu_{\rm obs}}^{X} < 0$ and $U_{\nu_{\rm obs}}^{X} > 0$, and to $\chi^{X} - \pi/2$ if $Q_{\nu_{\rm obs}}^{X} < 0$ and $U_{\nu_{\rm obs}}^{X} < 0$ \citep[e.g.,][]{Lan2018}.

\section{Results}

\begin{figure*}
\centering
\includegraphics[width=0.46\linewidth]{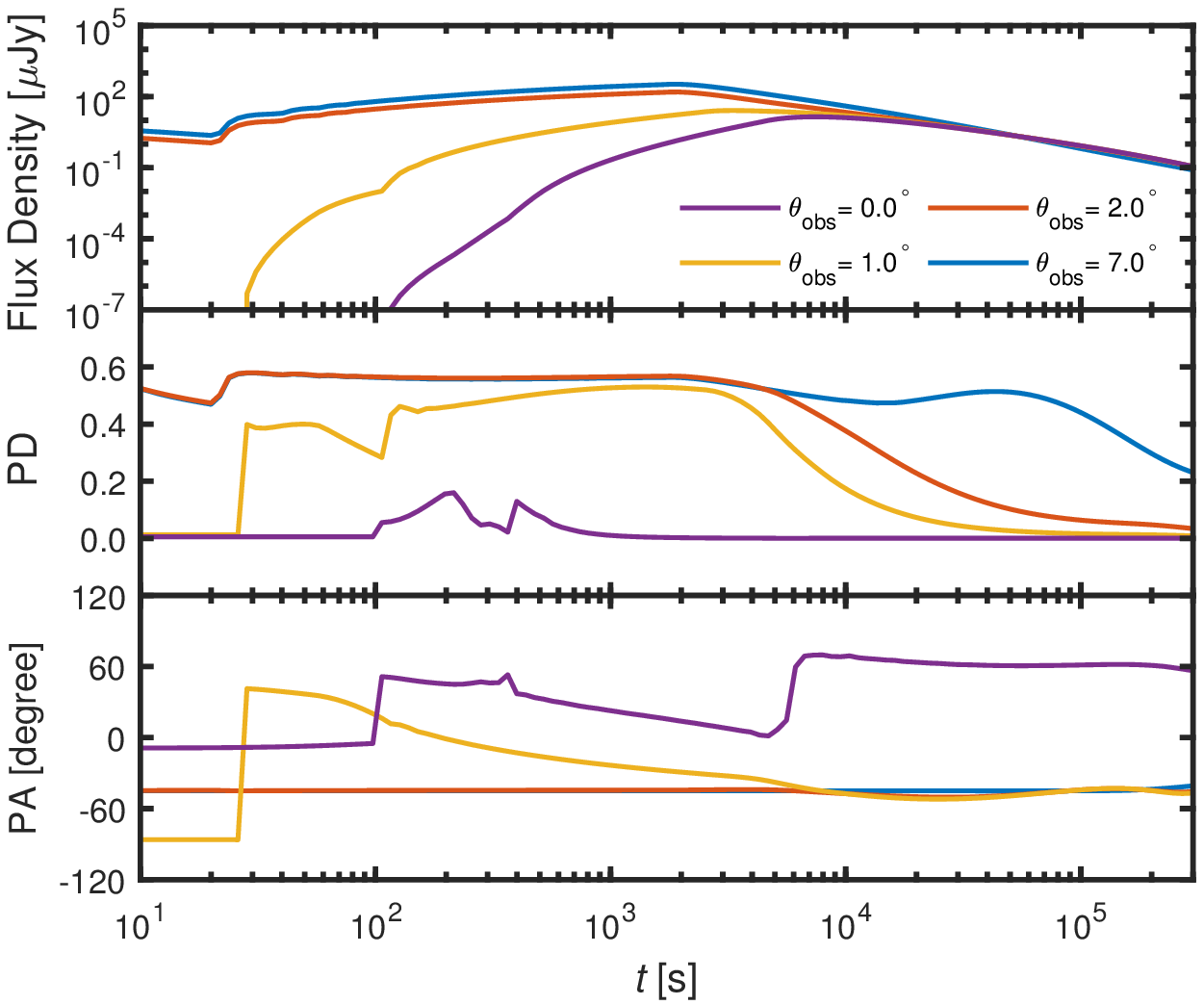}
\quad
\includegraphics[width=0.46\linewidth]{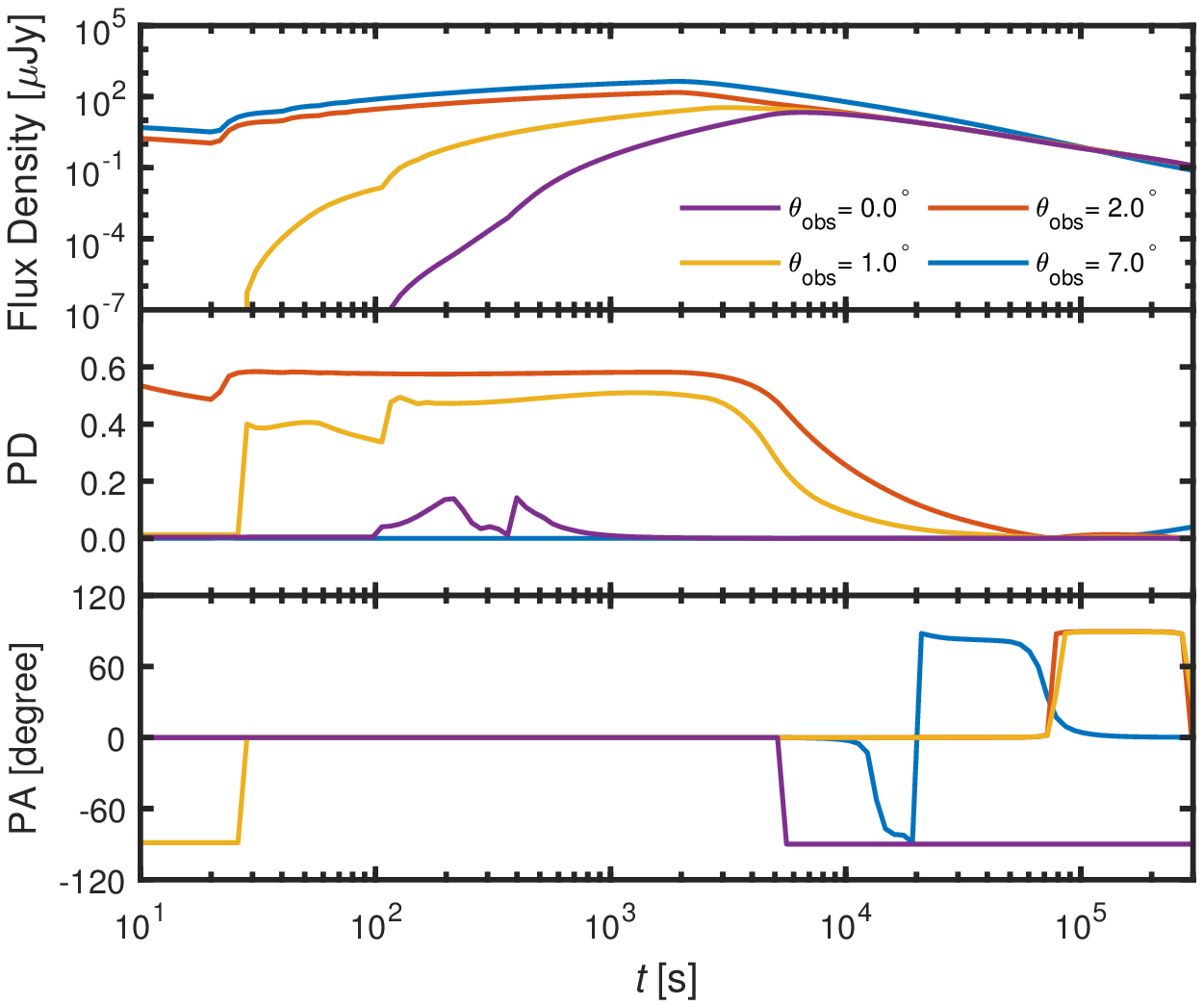}
\caption{Light curves and polarization evolution of GRB early optical afterglows in Case I. The left and right panels show the results with aligned and toroidal magnetic fields in Region 3, respectively, where Region 3 corresponds to the RS emission. The upper, middle, and bottom panels exhibit the light curves and the time evolution of the PD and PA, respectively. The violet, yellow, magenta, and blue lines denote the results obtained for $\theta_{\rm obs}=0^{\circ}$, $1^{\circ}$, $2^{\circ}$, and $7^{\circ}$, respectively.}
\label{fig2}
\end{figure*}

We numerically calculate the net linear polarization of the optical afterglows by considering the total contributions of the FS and RS in Regions 2 and 3, respectively, in two cases, i.e., the ordered magnetic field dominates (Case I) or the random magnetic field dominates (Case II). In our calculations, there is merely a slight dependency of the results across optical wavelengths as shown in \cite{Shimoda2021}, so we only take account of the R-band. We assume a top-hat jet with the half-opening angle $\theta_{\rm j} = 5^{\circ}$, consider the equal arrival time surface effect, and ignore the lateral expansion effect \citep[e.g.,][]{Zhang2009,Huang2021}. The age $t_{\rm end}$ of the jet beginning at $\phi_{\rm p}=180^{\circ}$ is set as $1 \, {\rm ks}$ (according to the activities of the central engines of LGRBs are able to arrive within thousands of seconds), and the precession period $T$ is set as $10 \, {\rm s}$ to satisfy the condition that the precession period is much less than the active timescale of the central engine. Note that $t_{\rm end}$ and $T$ are measured in the lab frame. Based on the expected value of the precession angle obtained by \cite{Stone2013}, we set the precession angle, i.e., $\theta_{\rm p} = 7^{\circ}$. For simplicity, we assume the same values for the dynamical parameters in both cases, including $L_{\rm ej} = 10^{49} \, {\rm erg~s^{-1}}$, $\Gamma_{\rm ej} = 300$, and $n_{1} = 1 \, {\rm cm^{-3}}$, except $\sigma_{\rm ej} = 10^{-1}$ in Case I and $10^{-3}$ in Case II. Meanwhile, the emission parameter values are as follows: $\varepsilon_{\rm e,2}=10^{-1}$, $\varepsilon_{B,2} = 10^{-3}$, and $p_{2} = p_{3} = 2.5$ in both cases, excluding $\varepsilon_{\rm e,3} = 10^{-1}$ in Case I and $10^{-3}$ in Case II, where $\varepsilon_{\rm e,2}$ ($\varepsilon_{\rm e,3}$) is the fraction of the internal energy density in the FS (RS) downstream that goes into the electrons and $p_{2}$ ($p_{3}$) is the power-law index of the energy distribution of the shock-accelerated electrons in the FS (RS) downstream. Moreover, here, we fix the linear PD $\Pi_{\rm 0} = 0.6 $ in the ordered magnetic field, the orientation of the aligned magnetic field $\delta = \pi/4$, and the typical redshift $z=1$. All parameters can be found in Table \ref{lab1}.

\subsection{Case I}

\begin{figure*}
\centering
\includegraphics[width=0.46\linewidth]{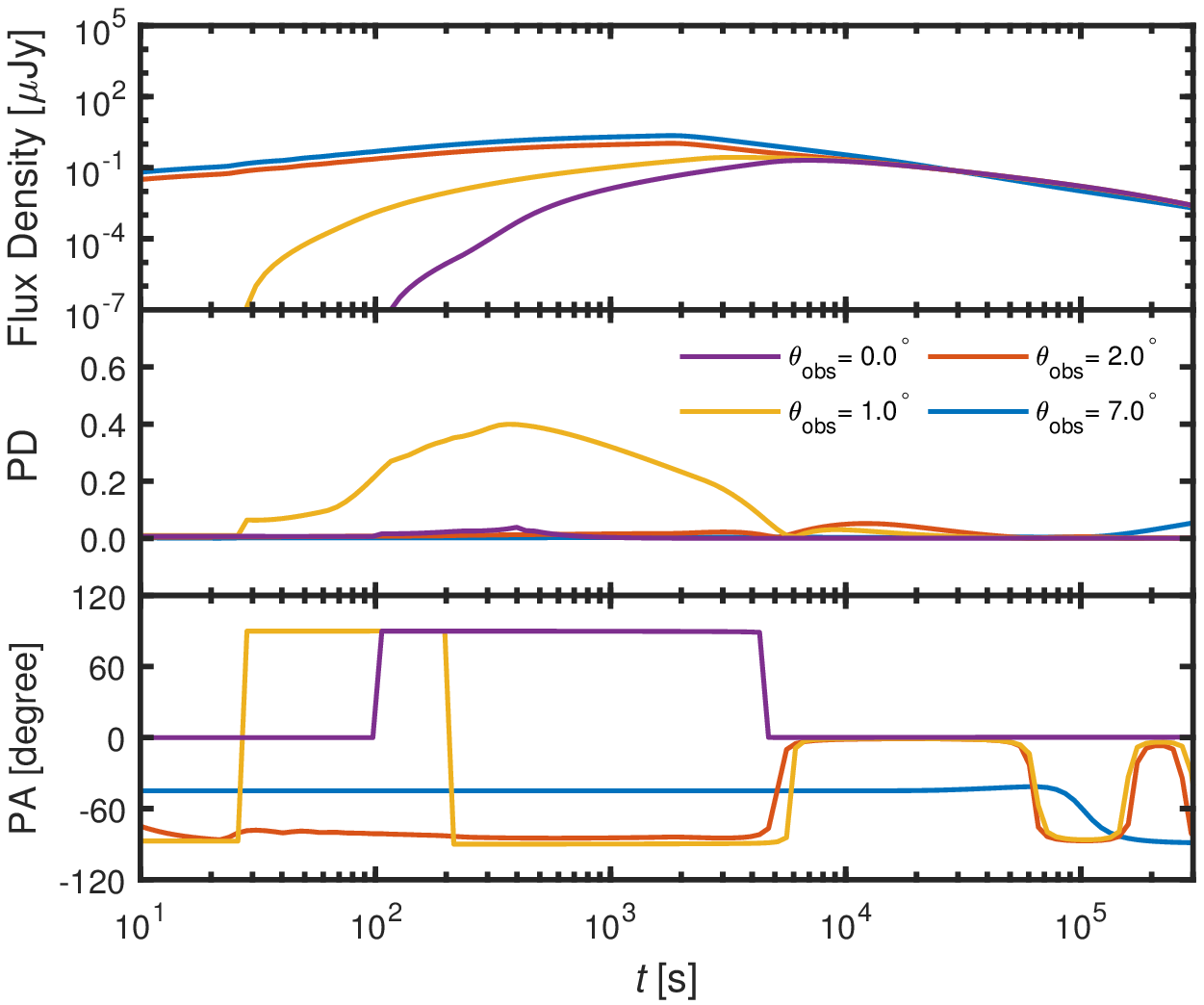}
\quad
\includegraphics[width=0.46\linewidth]{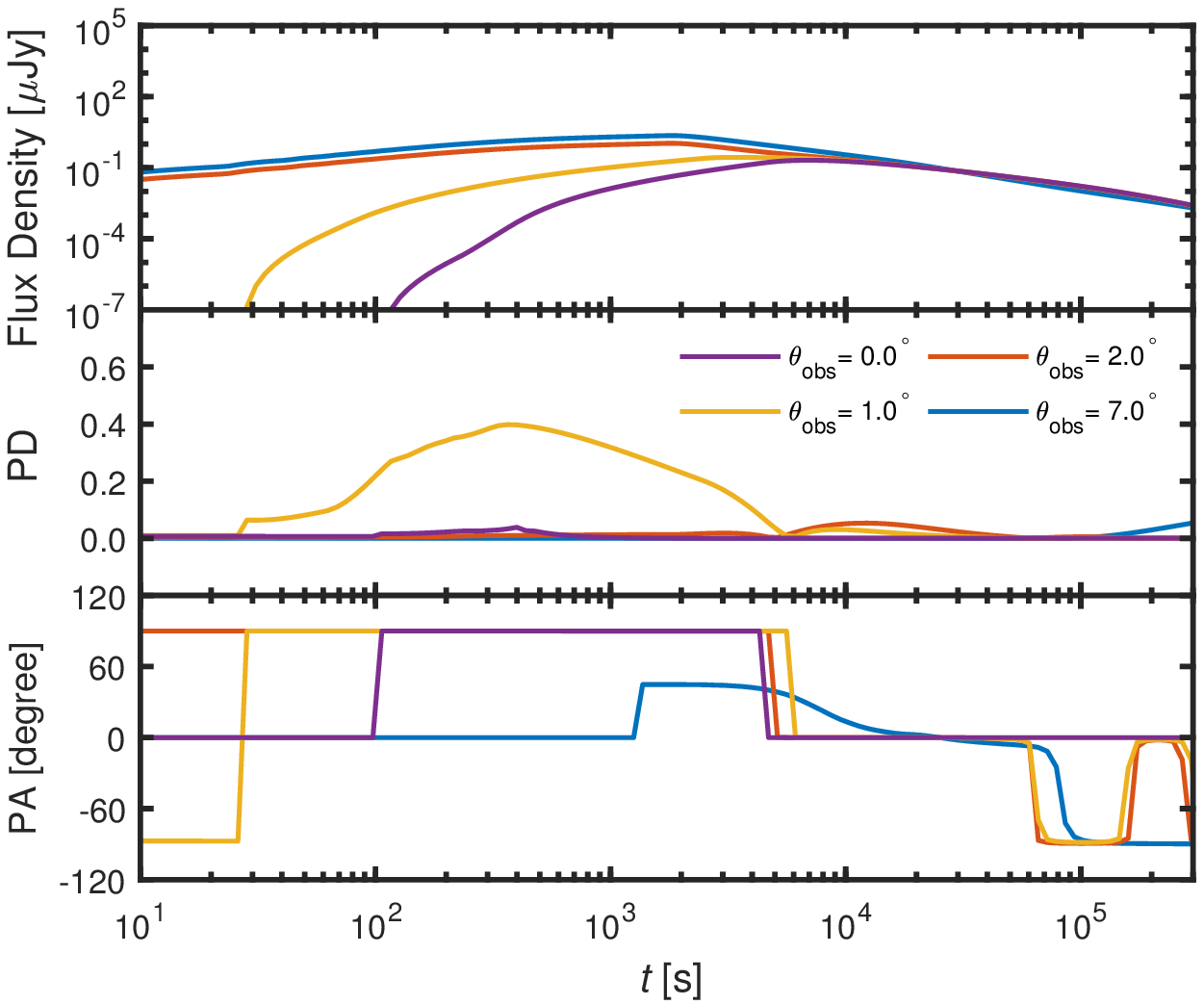}
\caption{Light curves and polarization evolution of GRB early optical afterglows that are the same as those in Figure \ref{fig2} but for Case II.}
\label{fig3}
\end{figure*}

Figure \ref{fig2} shows the results for Case I (the ordered magnetic field dominates). Noted that the ordered magnetic field all exists in Region 3, which corresponds to the RS emission. The left and right panels correspond to the aligned and toroidal magnetic field configurations in Region 3, respectively \citep{Spruit2001}. The upper, middle, and bottom panels exhibit the light curves and the time evolution of the PD and PA, respectively. The violet, yellow, magenta, and blue lines denote the results obtained for $\theta_{\rm obs}=0^{\circ}$, $1^{\circ}$, $2^{\circ}$, and $7^{\circ}$, respectively. Here, $\theta_{\rm obs}$ is the angle between the precession axis and the line of sight, as shown in Figure \ref{cartoon}.

The magnetic field topology has little influence in shaping the optical light curves (see the upper panels of Figure \ref{fig2}), whose emission approximately peaks at thousands of seconds post-burst. The hardly low fluxes for the light curves with $\theta_{\rm obs}= 0^{\circ}$ and $1^{\circ}$ in the early time are due to the delayed arrivals of the synchrotron photons as the line of sight deviates from the path of the jet precession. The light curves with $\theta_{\rm obs}= 2^{\circ}$ and $7^{\circ}$ have significantly higher fluxes in the early time because the line of sight points to the path of jet precession for $\theta_{\rm obs}= 2^{\circ}$ and $7^{\circ}$, which correspond the cases of the lines of sight on the edge and inside of the subjets, respectively. Additionally, the peak fluxes with $\theta_{\rm obs}= 0^{\circ}$ are approximately one order of magnitude lower than those with $\theta_{\rm obs}= 7^{\circ}$ due to the assumption of a top-hat jet for which the peak fluxes sharply decline when the line of sight moves from the inside to the outside of the precessing jets. This implies that the precession angle should not be too large, otherwise, a much lower flux will be undetectable \citep{Laskar2015}.

For the cases with $\theta_{\rm obs}=0^{\circ}$, the PDs (middle panels) are almost zero, regardless of the magnetic field configurations, which indicates that the high polarization produced in the ordered magnetic field can be almost counteracted in the precession process. However, for other values of $\theta_{\rm obs}$, high PDs appear. Once $\theta_{\rm obs}=7^{\circ}$, the PD reaches approximately $60\%$ in the aligned magnetic field configuration, whereas the value of the PD is close to zero in the toroidal configuration \citep{Lyutikov2003}. These results suggest that the effect of jet precession on the PD strongly depends on the position of the line of sight, namely, it is obvious only when the line of sight is around the precession axis. Here, the irregular shapes of the evolution profiles of the PD in the early time should arise from the assumption made for the dynamics, but these have no significant effect on the overall evolutionary trend of the PD.

For the PA (bottom panels), its evolution is characterized by basically constant, gradual decay, and abrupt changes. First, the constant evolution, such as that presented in the results obtained for $\theta_{\rm obs}=7^{\circ}$ with the aligned magnetic field configuration, is attributed to the invariant direction of the ordered magnetic field in the visible core (i.e., the $1/\Gamma$ core). Second, the gradual decay, in the situation for $\theta_{\rm obs}=1^{\circ}$ with the aligned magnetic field configuration, occurs when the direction of the ordered magnetic field in the visible core gradually changes due to the increase in the observational regions of the subjets moving in different positions with the decrease Lorentz factors. Third, the appearance of the abrupt changes, including both the abrupt changes for $90^{\circ}$ and $180^{\circ}$, depends on the transition of the signs of $Q_{\nu_{\rm obs}}^{X}$ or $U_{\nu_{\rm obs}}^{X}$, reflecting the abrupt changes in the direction of the ordered magnetic field. When the sign of $Q_{\nu_{\rm obs}}^{X}$ changes, abrupt changes in $90^{\circ}$ occur. In contrast, when the sign of $U_{\nu_{\rm obs}}^{X}$ changes with $Q_{\nu_{\rm obs}}^{X}<0$, abrupt changes in $180^{\circ}$ appear, as shown in the results obtained for $\theta_{\rm obs}= 7^{\circ}$ with the toroidal magnetic field configuration. Therefore, which of these features appears mainly depends on how the direction of the ordered magnetic field in the visible core changes.

\subsection{Case II}

Figure \ref{fig3} shows the results obtained for Case II (the random magnetic field dominates). Here, Case II corresponds the dominance of the FS emission because the random magnetic field exists in Region 2. The left and right panels still correspond to the aligned and toroidal magnetic field configurations in Region 3, respectively.

One can find that the light curves and the evolutions of the PD in both panels maintain consistency. Meanwhile, the values of the PDs are significantly low, except in the situation of $\theta_{\rm obs}=1^{\circ}$. In the situation of $\theta_{\rm obs}=0^{\circ}$, the PDs are almost zero, which is due to the jet precession.

The evolution of PAs appears to be more complex. Most of the evolution profiles exhibit irregularly abrupt changes over two times, which suggests that the direction of the total magnetic fields is irregularly changeable. This should be attributed to the irregular changes in the geometry of the emission region in the visible core with a decrease in the Lorentz factors. In addition, since the signs of $Q_{\nu_{\rm obs}}^{X}$ and $U_{\nu_{\rm obs}}^{X}$ with significantly small values are sensitive to the mathematical definition, related features for the PAs should be treated carefully.

\section{Summary}

It is expected that jet precessions exist in GRBs, and polarimetry plays an important role in the study of GRBs. These results motivate interest in determining whether the jet precession can significantly affect the polarization behavior, especially in the early optical afterglows. Consequently, we investigate the polarization evolution of early optical afterglows in the framework of long-lived precessing jets with either dominant ordered magnetic field or random component (i.e. scenarios with mixed properties). As a result, we find that the high levels of PDs expected in the RS emission, due to the large-scale magnetic fields ejected from the central engine, can be significantly reduced by the jet precession.

In our model, the depolarizing effect induced by the jet precession still exists for the cases with other different active timescales of the central engines and the precession periods. For the case with more precession laps in an engine active timescale, the depolarizing effect might be more significant. Moreover, the decreasing precession angle should increase the polarizing effect.

Recently, GRB 190114C, the first TeV-detected GRB \citep{MAGIC2019}, is reported to have a low PD, $\sim 2\%-7\%$, in its early afterglows, and the low PD can be well explained by the large-scale magnetic fields distorted before the RS emission \citep[][and references therein]{Jordana2020}. Actually, the low PD of the early afterglow of GRB 190114C can also be interpreted by the depolarizing effect in jet precession model.

In addition to the early afterglows, low PDs can be detected in the prompt $\gamma$-ray emission \citep{Zhang2019,Kole2020}. \cite{Zhang2019} reported a sample of polarization measurements obtained using the dedicated GRB polarimeter (POLAR) and obtained an average PD, $\sim 10\%$, which is lower than that predicted by some popular models. \cite{Lan2021} proposed that these results can be explained by a model involving a mixed magnetic field (i.e., by entangling the large-scale ordered component present in the ejecta). However, the origin of the mixed magnetic fields is still unclear. Inspired by our results, we suspect that jet precession is a plausible mechanism for building mixed magnetic fields. For example, in the framework of the massive collapsars, due to the high accretion rate and the redistribution of the fallback debris during the first few to tens of seconds, the precession of the jets should be quite fast, violent and even irregular \citep{Liu2018,Huang2021}. This will lead to a distortion of the large-scale ordered magnetic fields.

\acknowledgments
We thank the anonymous referee for helpful suggestions and comments. This work was supported by the National Natural Science Foundation of China under grants 12173031 and 11822304, and the science research grants from the China Manned Space Project with No. CMS-CSST-2021-B11.

\end{document}